\documentclass[final,3p,times,twocolumn]{elsarticle}
\usepackage{epsfig,bm}
\usepackage{graphics}
\usepackage{amsmath}
\usepackage{mathrsfs}
\usepackage[normalem]{ulem}  
\usepackage[dvips]{color} 

\renewcommand\sout{\bgroup \color{red} \ULdepth=-.5ex \ULset}

\biboptions{sort&compress}

\begin{document}

\begin{frontmatter}

\title{Double ratio of charmonia in $p+\textrm{Pb}$\ collisions at $\sqrt{s_{\rm NN}}=5.02$\ TeV}

\author[ic]{Yunpeng Liu}
\author[ic,ph]{Che Ming Ko}
\author[ic,fr]{Taesoo Song}

\address[ic]{Cyclotron Institute, Texas A$\&$M University, College Station, Texas 77843, USA}
\address[ph]{Department of Physics and Astronomy, Texas A$\&$M University, College Station, Texas 77843, USA}%
\address[fr]{Frankfurt Institute for Advanced Studies, J. W. Goethe University, 60438 Frankfurt am Main, Germany}%
\begin{abstract}
Based on a kinetic description of $J/\psi$ dissociation and production in  an expanding quark-gluon plasma  that is described by a 2+1 dimensional ideal hydrodynamics, we have studied the double ratio $R_{p\textrm{Pb}}(\psi')/R_{p\textrm{Pb}}(J/\psi)$ of charmonia in $p+\textrm{Pb}$ collisions at $\sqrt{s_{\rm NN}}=5.02$\ TeV by including not only the cold nuclear matter effects but also the hot nuclear matter effects. We find that the double ratio of prompt charmonia is significantly suppressed in the most central collisions  as a result of the hot nuclear matter effects.
\end{abstract}


\end{frontmatter}

Charmonia suppression in heavy ion collisions  was first suggested in 
Ref.~\cite{Matsui:1986dk} based on the  consideration of color screening in quark-gluon plasma (QGP). The  anomalous suppression of $J/\psi$ production, besides the normal 
suppression due to the cold nuclear matter effects, observed in Pb+Pb collisions at the Super Proton Synchrotron (SPS) seemed to confirm  this prediction and thus suggested that a 
QGP was produced in these collisions~\cite{Gonin:1996wn}. Since then, there have been extensive 
experimental and theoretical studies on $J/\psi$ production and suppression  in both elementary and nuclear collisions. In particular, 
\begin{itemize}
	\item  Baseline study of charmonia production in $p+p$ collisions has been carried out using the  color singlet model (CSM)~\cite{Chang:1979nn} and/or the color octet model (COM)~\cite{Bodwin:1994jh}.
	\item  Cold nuclear matter effects,  including the shadowing effect, the Cronin effect and the nuclear absorption, have been studied in $p$+A and $d$+A collisions.
	\item  Hot medium effects,  including the color screening, gluon dissociation, quasi-free scattering dissociation, regeneration and so forth, have been studied in $A+A$ collisions.
\end{itemize}
However, with the high beam energy at the LHC,  the initial energy density  in $p$+Pb collisions can be sufficient large for the formation of a QGP.  It is then of interest to study the hot  medium effects  on quarkonia production in these collisions and compare them to those due to  the cold nuclear matter~\cite{Gerschel:1988wn, Gousset:1996xt, McLerran:1998nk, Kopeliovich:1991pu, Kopeliovich:1993pw, Kopeliovich:2003tz, Iancu:2003xm, Ferreiro:2008wc}.

In Ref.~\cite{Liu:2013via}, we have carried out such a study by using a 2+1 dimensional ideal hydrodynamic model to describe the  bulk dynamics of $p+\textrm{Pb}$ collisions at $\sqrt{s_{\textrm{NN}}}=5.02$ TeV, i.e., based on the  equation 
\begin{eqnarray}
   \partial_{\mu}T^{\mu\nu}&=& 0
\end{eqnarray}
with the boost invariant condition. In the above,  $T^{\mu\nu}=(\epsilon+p)u^{\mu}u^{\nu}-g^{\mu\nu}p$\ is the energy-momentum tensor in terms of the energy density $\epsilon$, pressure $p$, and four velocity $u$.  For the equation of state $\epsilon(p)$, we have used that of Ref.~\cite{Liu:2009nb} based on an ideal gas of  massless partons for the QGP and massive  hadrons for the hadronic matter, together with a bag constant which leads to a critical temperature $T_c=165$ MeV~\cite{Aoki:2006br, Borsanyi:2010bp, Aoki:2006we}  at zero baryon  chemical potential. For the initial energy density at time $\tau_0=0.6$\ fm$/c$ and its rapidity dependence, it is obtained from a multiple phase transport (AMPT) model~\cite{Lin:2004en}.   Selecting the most central $10\%$ collisions  according to the yield of partons in AMPT, the maximum initial energy density is found at rapidity $y=-2$.

\begin{figure}[h]
    \centering
    {\includegraphics[width=0.23\textwidth]{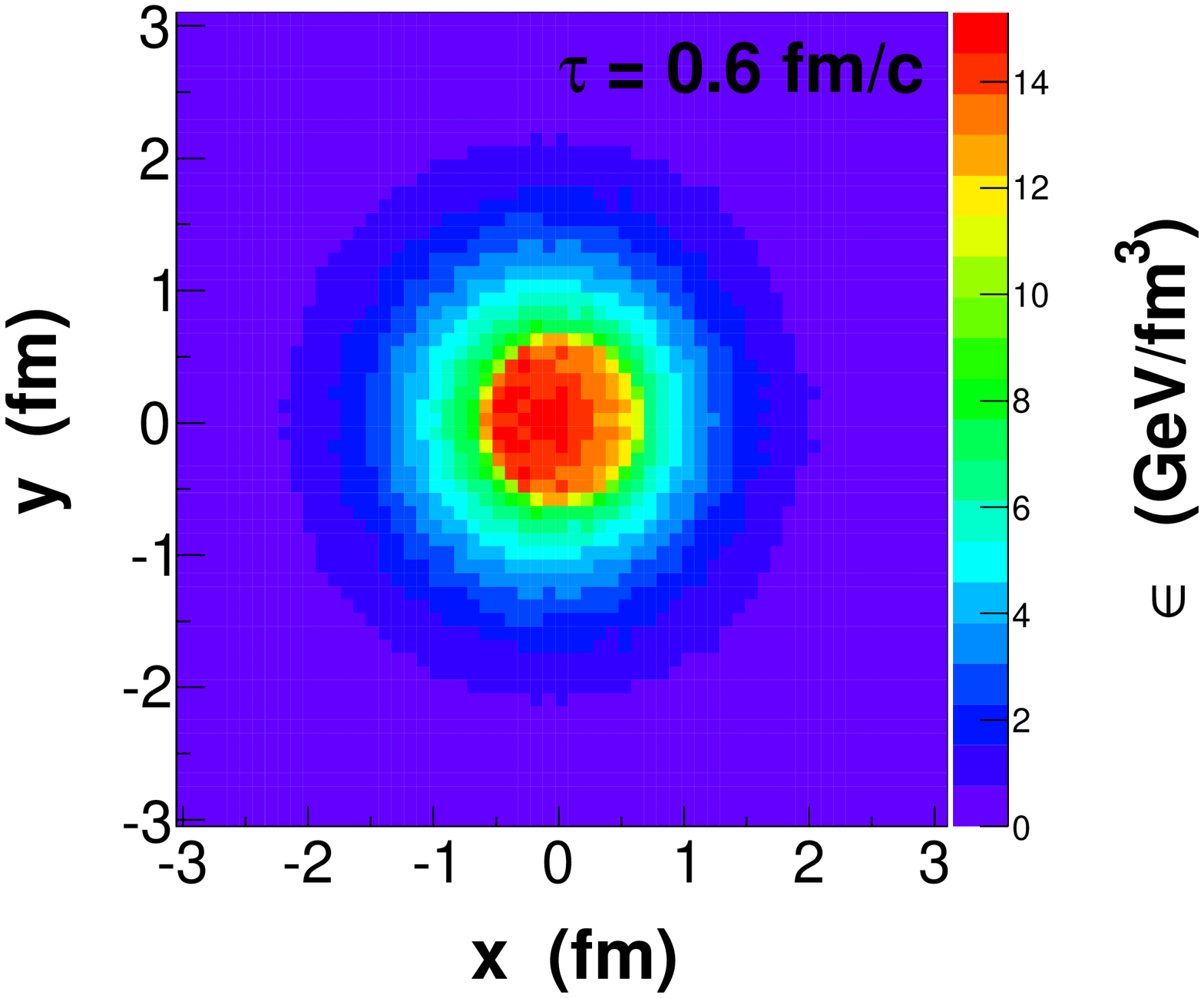}
    \includegraphics[width=0.23\textwidth]{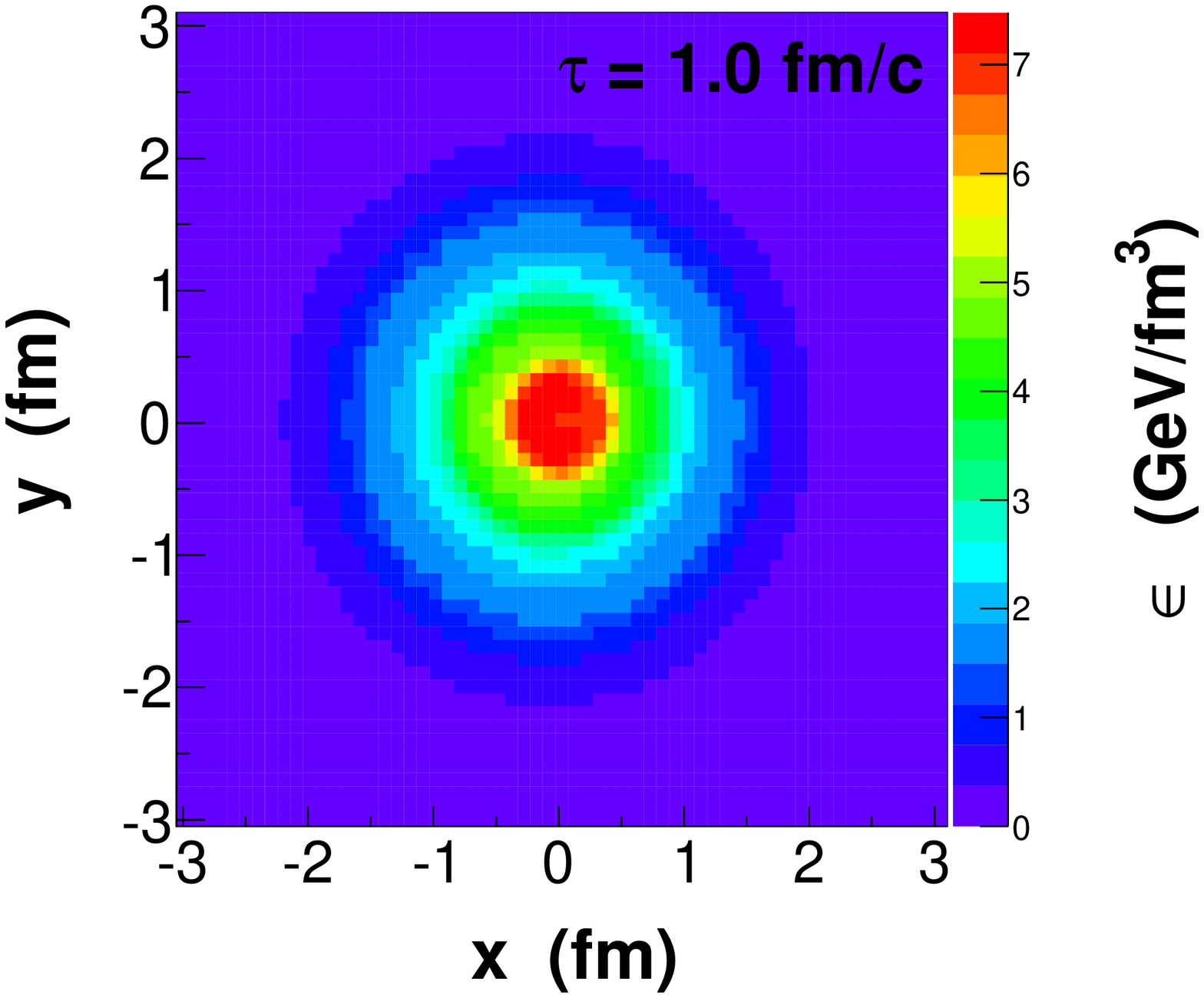}}
    {\includegraphics[width=0.23\textwidth]{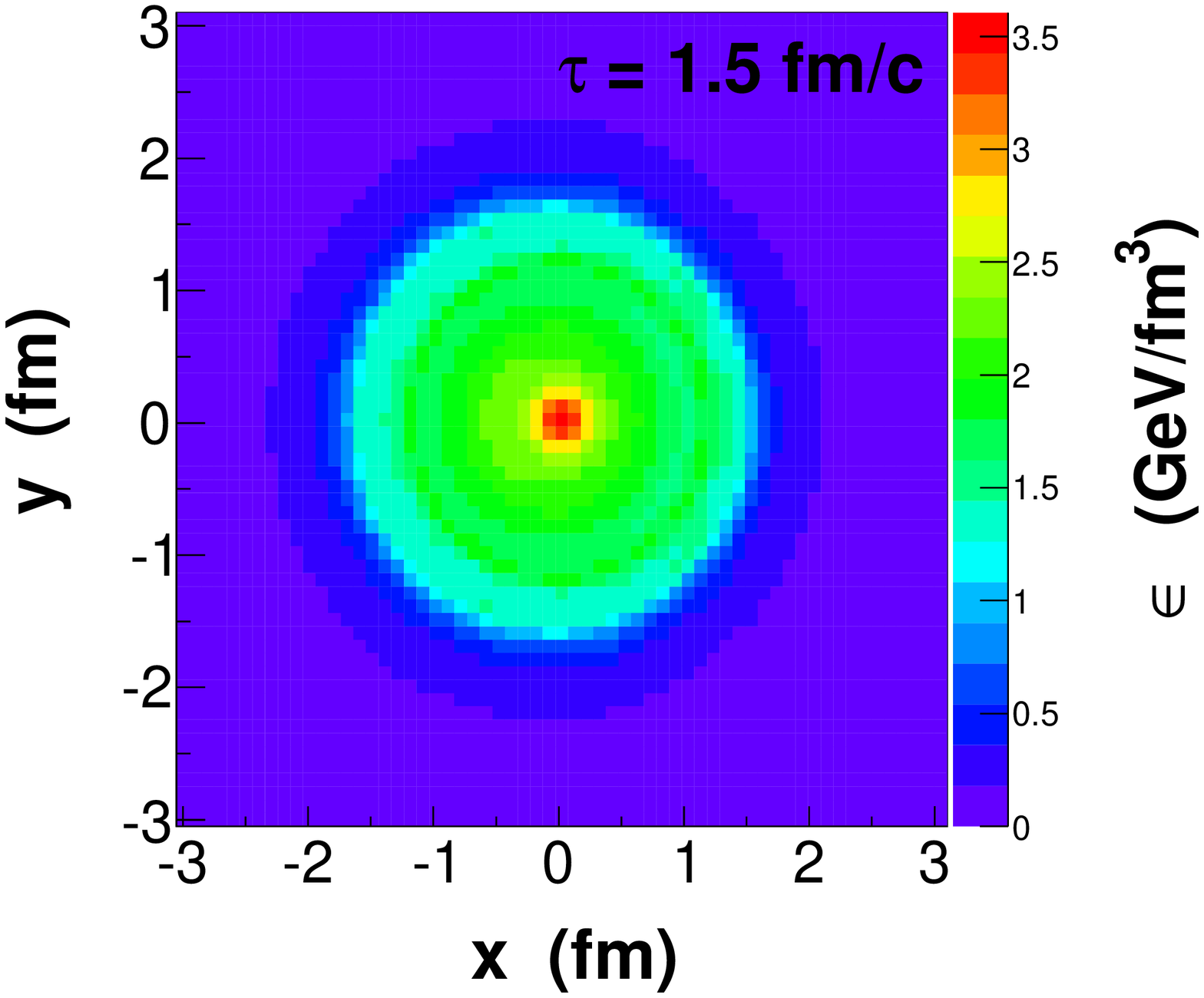}
    \includegraphics[width=0.23\textwidth]{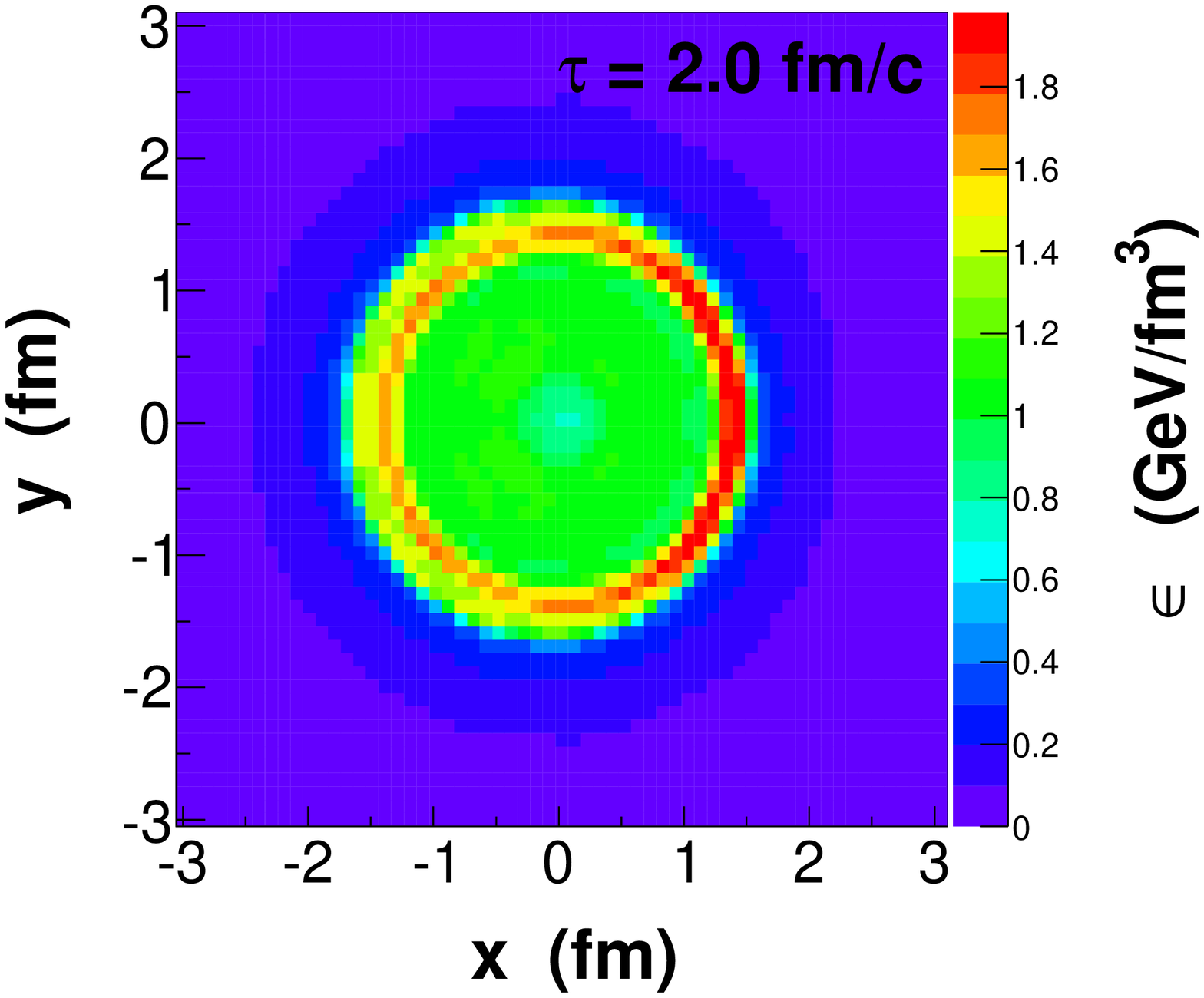}}
    \caption{(Color online) Energy density $\epsilon$ at rapidity $y=-2$ in the most central $10\%$  $p+\textrm{Pb}$ collisions at $\sqrt{s_{\rm NN}}=5.02$ TeV  as a function of the transverse coordinate $x$\ and $y$\ relative to the center of the proton at different proper time $\tau$.}
    \label{fig_epsilon_xy}
\end{figure}

In Fig.~\ref{fig_epsilon_xy}, we show the energy density distribution in the transverse plane at rapidity $y=-2$  for different proper time $\tau$. The origin of the coordinates is located at the center of the proton, and the center of the lead lies on the negative x-axis. The initial energy density is roughly isotropic in the transverse plane with  a slightly larger value on the lead side. The maximum value is $\epsilon_0\approx 15$\ GeV$/$fm$^3$, which corresponds to a temperature $T_0\approx 290$\ MeV.
\par
Since the energy density is large and the volume is small compared with  those in A+A collisions at lower energies, the system expands fast, leading  to a quick decreas of the energy density at the center. At time $\tau=2.0$\ fm$/c$, the energy density at the center is even smaller than that in the peripheral region, and the maximum energy density is at the radius $r\approx1.5$\ fm. The value of the energy density on the positive axis is slightly larger than that at the negative axis.  Since the center of the  QGP produced in $p$+Pb collisions hadronizes earlier,  its spacial topology  differs from that in Pb+Pb collisions\footnote{The Euler characteristic $\chi$ changes from $2$ to $4$ in central $p+$Pb collisions.}.

For the time evolution of the distribution function $f({\bf x}, {\bf p}, t)$ of $J/\psi$ in the phase space of coordinate ${\bf x}$ and momentum ${\bf p}$, we use  the transport equation
\begin{eqnarray}
   \partial_t f+{\bf v}\cdot\nabla f& =& -\alpha f + \beta,
   \label{eq_transport}
\end{eqnarray}
where ${\bf v}$\ is the velocity of the $J/\psi$, and $\alpha$\ and $\beta$\ are the  dissociation and the regeneration rate, respectively. The dissociation rate
\begin{eqnarray}
   \alpha(T,u,p)&=& \frac{N_g}{E}\int\frac{d{\bf k}}{(2\pi)^3E_g}{k}\cdot{p}f_g(T,u, k)\sigma_D.\nonumber\\
   \label{eq_alpha}
\end{eqnarray}
takes into account the gluon dissociation process $J/\psi+g\rightarrow c+\bar{c}$ with the cross section $\sigma_D$~\cite{Peskin:1979va,Manceau:2013zta,Liu:2009nb}. The regeneration rate takes into account the inverse process $c+\bar c\rightarrow J/\psi+g$, whose cross section is related to that of the gluon dissociation process by the detailed balance  relation, and can be calculated using the  distribution functions of charm and anticharm quarks. For simplicity, we take the charm quark distribution  to be in thermal equilibrium, i.e.,
\begin{eqnarray}
   f_c({\bf x}, {\bf q}_c)&=& \rho_c({\bf x})\frac{N}{e^{q_c\cdot u/T}+1},
\end{eqnarray}
where $\rho_c$\ is the number density of charm quarks and satisfies the conservation equation $\partial_{t}\rho_c+\nabla\cdot(\rho_c{\bf v}_m)=0$ with ${\bf v}_m$\ being the velocity of the medium, and $N\equiv \left[(2\pi)^{-3}\int d {\bf q}_c\left(e^{q_c\cdot u/T}+1\right)^{-1}\right]^{-1}$\ is the normalization factor for the Fermi distribution.  For  anticharm quarks, their distribution function  $f_{\bar c}$ has a similar form. We also include a velocity-dependent dissociation temperature $T_d$  to describe the color screening effect~\cite{Liu:2012zw}.
\par
Different from  Pb+Pb collisions at the LHC,  at most one pair of charm and anticharm quarks 
are produced in  a $p$+Pb collision, and they are  also not likely to reach thermal equilibrium . These effects are included via a  canonical enhancement factor $C_{ce}=1+1/(dN_c^{\rm dir}/dy)$ in the evaluation of the equilibrium number of $J/\psi$, as in  the statistical model~\cite{Andronic:2006ky, Kostyuk:2005zd, Liu:2012tn}, and the relaxation reduction factor $r=1-\exp(-\tau/\tau_r)$~\cite{Grandchamp:2002wp} with the charm quark relaxation time $\tau_r=7$ fm/$c$~\cite{Zhao:2007hh, Song:2012at}. More details can be found in Ref.~\cite{Liu:2013via}. We note that these  effects can be  more accurately studied in  a transport model~\cite{Liu:2014rsa}.
\par
For the initial distribution of $J/\psi$, it is obtained from the Glauber model with the inclusion of initial-state cold nuclear matter effects and  a $J/\psi$ production cross section of  $d\sigma_{pp}/dy=5.68$\ $\mu$b for $pp$ collisions at $5.02$ TeV. The latter is  obtained from interpolating the experimental results at lower ($\sqrt{s_{\textrm{NN}}}=2.76$ TeV)~\cite{Abelev:2012kr} and higher ($\sqrt{s_{\textrm{NN}}}=7$ TeV)~\cite{Aamodt:2011gj} energies with a power-law form  and an average transverse momentum square $\langle p_T^2\rangle=9.5$ GeV$^2$ at middle rapidity.  As to the rapidity distribution, it is assumed to be Gaussian with parameters  taken from Refs.~\cite{Bossu:2011qe, Liu:2009wza}. The density distribution of initially produced $J/\psi$ in space is assumed to be proportional to the thickness of a uniform solid  sphere of radius $r=0.8$ fm as that for the proton.  
\par
 For the cold nuclear matter effects  on $J/\psi$\ and charm quarks, they are assumed to be the same  and are taken from Refs.~\cite{Albacete:2013ei,  Manceau:2013zta},  which then lead to a suppression in proton rapidity and an enhancement in Pb rapidity. 
\par
The above treatment for the ground state $J/\psi$  is generalized to include its excited states $\chi_c$ and $\psi^\prime$ with different dissociation temperatures $T_D$ and cross sections~\cite{Wang:2002ck, Arleo:2001mp} as in Ref.~\cite{Liu:2012zw}. For the initial abundance of these excited states, they are determined from that of $J/\psi$ by using the empirically known feed-down contributions to $J/\psi$ production in   $p+p$ collisions,  i.e., using the proportion $6:3:1$ for  direct $J/\psi$, feed-down from $\chi_c$ and  from $\psi^\prime$~\cite{Zoccoli:2005yn}. They are further assumed to  subject to the same cold nuclear matter effects as the $J/\psi$.
 
For the contribution to $J/\psi$ production from the decay of regenerated $\chi_c$ and 
$\psi^\prime$, they are included with the branch ratios from Ref.~\cite{Beringer:1900zz}. 
  
\begin{figure}[hbt]
    \centering
    \includegraphics[width=0.48\textwidth]{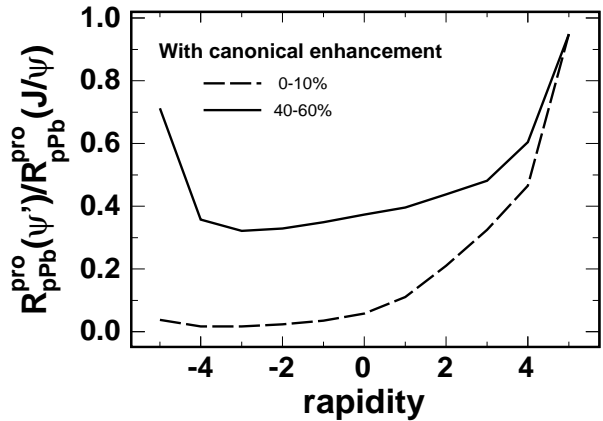}
    \caption{Double ratio $R_{p{\rm Pb}}^{\textrm{pro}}(\psi')/R_{p{\rm Pb}}^{\textrm{pro}}(J/\psi)$ of prompt charmonia in $p+\textrm{Pb}$ collisions at $\sqrt{s_{\textrm{NN}}}=5.02$\ TeV as a function of rapidity for centralities $0\textrm{-}10\%$\ and $40\textrm{-}60\%$.} 
    \label{fig_raa}
\end{figure}

In Fig.~\ref{fig_raa}, we show the double ratio of prompt charmonia as a function of rapidity. It  is seen that the double ratio at rapidity from $-4$ to $0$ is smaller than unit due to stronger suppression of $\psi'$\ in the hot medium. In the most central $10\%$\ collisions, it is smaller than $0.1$, while in  collisions with centrality $40\%\textrm{-}60\%$, it is between $0.3$\ and $0.4$. Since the formation time of $J/\psi$\ and $\psi'$\ is  much longer than that for the proton and Pb to pass through each other, the cold nuclear matter effects are not expected to differ much between the $J/\psi$\ and $\psi'$.

\begin{figure}[hbt]
     \centering
     \includegraphics[width=0.48\textwidth]{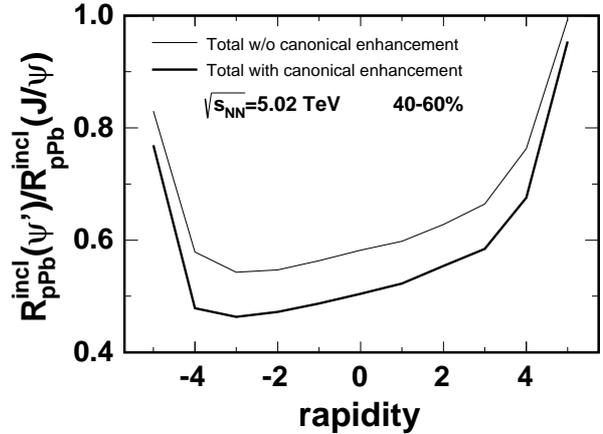}
     \caption{Double ratio $R_{p{\rm Pb}}^{\textrm{incl}}(\psi')/R_{p{\rm Pb}}^{\textrm{incl}}(J/\psi)$ of inclusive charmonia in $p+\textrm{Pb}$ collisions at $\sqrt{s_{\textrm{NN}}}=5.02$\ TeV as a function of rapidity for centrality $40\textrm{-}60\%$.} 
     \label{fig_dr}
\end{figure}

In experiments, the inclusive double ratio is easier to measure. According to a recent study~\cite{Chen:2013wmr}, the contribution of $B$\ meson decay increases the double ratio in heavy ion collisions. The corresponding result for $p$+Pb collisions is shown in Fig.~\ref{fig_dr}. The qualitative effect of $B$ meson decay is the same as that in Pb+Pb collisions~\cite{Chen:2013wmr}, and the double ratio is above $0.4$\ in the whole rapidity range.

In summary, we  have studied the double ratio of quarkonia in $p+\textrm{Pb}$ collisions at $\sqrt{s_{\rm NN}}=5.02$ TeV in a transport approach based on a 2+1 dimensional ideal hydrodynamic model and found that both the prompt and inclusive double ratios  are smaller than  one, especially for the prompt one in the most central $10\%$ collisions.  This is not expected  from the cold nuclear matter effects as they  exist before  $J/\psi$\ and $\psi'$ are formed.  Measuring the double ratio of charmonia thus provides the possibility to  study the hot  medium effects in $p+$Pb collisions at the LHC.

\section*{Acknowledgements}
 
This work was supported by the U.S. National Science Foundation under Grant No. PHY-1068572, the US Department of Energy under Contract No. DE-FG02-10ER41682, and the Welch Foundation under Grant No. A-1358.
\bibliographystyle{elsarticle-num}
\bibliography{ref}

\begin{thebibliography}{10}
\expandafter\ifx\csname url\endcsname\relax
  \def\url#1{\texttt{#1}}\fi
\expandafter\ifx\csname urlprefix\endcsname\relax\def\urlprefix{URL }\fi
\expandafter\ifx\csname href\endcsname\relax
  \def\href#1#2{#2} \def\path#1{#1}\fi

\bibitem{Matsui:1986dk}
T.~Matsui, H.~Satz, {J/psi Suppression by Quark-Gluon Plasma Formation}, Phys.
  Lett. B178 (1986) 416.
\newblock \href {http://dx.doi.org/10.1016/0370-2693(86)91404-8}
  {\path{doi:10.1016/0370-2693(86)91404-8}}.

\bibitem{Gonin:1996wn}
M.~Gonin, et~al., {Anomalous J / psi suppression in Pb + Pb collisions at
  158-A-GeV/c}, Nucl. Phys. A610 (1996) 404c--417c.
\newblock \href {http://dx.doi.org/10.1016/S0375-9474(96)00373-9}
  {\path{doi:10.1016/S0375-9474(96)00373-9}}.

\bibitem{Chang:1979nn}
C.-H. Chang, {Hadronic Production of $J/\psi$ Associated With a Gluon}, Nucl.
  Phys. B172 (1980) 425--434.
\newblock \href {http://dx.doi.org/10.1016/0550-3213(80)90175-3}
  {\path{doi:10.1016/0550-3213(80)90175-3}}.

\bibitem{Bodwin:1994jh}
G.~T. Bodwin, E.~Braaten, G.~P. Lepage, {Rigorous QCD analysis of inclusive
  annihilation and production of heavy quarkonium}, Phys. Rev. D51 (1995)
  1125--1171.
\newblock \href {http://arxiv.org/abs/hep-ph/9407339}
  {\path{arXiv:hep-ph/9407339}}, \href
  {http://dx.doi.org/10.1103/PhysRevD.55.5853, 10.1103/PhysRevD.51.1125}
  {\path{doi:10.1103/PhysRevD.55.5853, 10.1103/PhysRevD.51.1125}}.

\bibitem{Gerschel:1988wn}
C.~Gerschel, J.~Hufner, {A Contribution to the Suppression of the J/psi Meson
  Produced in High-Energy Nucleus Nucleus Collisions}, Phys. Lett. B207 (1988)
  253--256.
\newblock \href {http://dx.doi.org/10.1016/0370-2693(88)90570-9}
  {\path{doi:10.1016/0370-2693(88)90570-9}}.

\bibitem{Gousset:1996xt}
T.~Gousset, H.~Pirner, {The Ratio of gluon distributions in Sn and C}, Phys.
  Lett. B375 (1996) 349--354.
\newblock \href {http://arxiv.org/abs/hep-ph/9601242}
  {\path{arXiv:hep-ph/9601242}}, \href
  {http://dx.doi.org/10.1016/0370-2693(96)00189-X}
  {\path{doi:10.1016/0370-2693(96)00189-X}}.

\bibitem{McLerran:1998nk}
L.~D. McLerran, R.~Venugopalan, {Fock space distributions, structure functions,
  higher twists and small x}, Phys. Rev. D59 (1999) 094002.
\newblock \href {http://arxiv.org/abs/hep-ph/9809427}
  {\path{arXiv:hep-ph/9809427}}, \href
  {http://dx.doi.org/10.1103/PhysRevD.59.094002}
  {\path{doi:10.1103/PhysRevD.59.094002}}.

\bibitem{Kopeliovich:1991pu}
B.~Kopeliovich, B.~Zakharov, {Quantum effects and color transparency in
  charmonium photoproduction on nuclei}, Phys. Rev. D44 (1991) 3466--3472.
\newblock \href {http://dx.doi.org/10.1103/PhysRevD.44.3466}
  {\path{doi:10.1103/PhysRevD.44.3466}}.

\bibitem{Kopeliovich:1993pw}
B.~Kopeliovich, J.~Nemchick, N.~N. Nikolaev, B.~Zakharov, {Decisive test of
  color transparency in exclusive electroproduction of vector mesons}, Phys.
  Lett. B324 (1994) 469--476.
\newblock \href {http://arxiv.org/abs/hep-ph/9311237}
  {\path{arXiv:hep-ph/9311237}}, \href
  {http://dx.doi.org/10.1016/0370-2693(94)90225-9}
  {\path{doi:10.1016/0370-2693(94)90225-9}}.

\bibitem{Kopeliovich:2003tz}
B.~Kopeliovich, {Transparent nuclei and deuteron gold collisions at RHIC},
  Phys. Rev. C68 (2003) 044906.
\newblock \href {http://arxiv.org/abs/nucl-th/0306044}
  {\path{arXiv:nucl-th/0306044}}, \href
  {http://dx.doi.org/10.1103/PhysRevC.68.044906}
  {\path{doi:10.1103/PhysRevC.68.044906}}.

\bibitem{Iancu:2003xm}
E.~Iancu, R.~Venugopalan, {The Color glass condensate and high-energy
  scattering in QCD}\href {http://arxiv.org/abs/hep-ph/0303204}
  {\path{arXiv:hep-ph/0303204}}.

\bibitem{Ferreiro:2008wc}
E.~Ferreiro, F.~Fleuret, J.~Lansberg, A.~Rakotozafindrabe, {Cold nuclear matter
  effects on J/psi production: Intrinsic and extrinsic transverse momentum
  effects}, Phys. Lett. B680 (2009) 50--55.
\newblock \href {http://arxiv.org/abs/0809.4684} {\path{arXiv:0809.4684}},
  \href {http://dx.doi.org/10.1016/j.physletb.2009.07.076}
  {\path{doi:10.1016/j.physletb.2009.07.076}}.

\bibitem{Liu:2013via}
Y.~Liu, C.~M. Ko, T.~Song, {Hot medium effects on $J/\psi$ production in $p+Pb$
  collisions at $\sqrt{s_{NN}}=5.02$ TeV}, Phys. Lett. B728 (2014) 437--442.
\newblock \href {http://arxiv.org/abs/1309.5113} {\path{arXiv:1309.5113}},
  \href {http://dx.doi.org/10.1016/j.physletb.2013.12.016}
  {\path{doi:10.1016/j.physletb.2013.12.016}}.

\bibitem{Liu:2009nb}
Y.-p. Liu, Z.~Qu, N.~Xu, P.-f. Zhuang, {J/psi Transverse Momentum Distribution
  in High Energy Nuclear Collisions at RHIC}, Phys. Lett. B678 (2009) 72--76.
\newblock \href {http://arxiv.org/abs/0901.2757} {\path{arXiv:0901.2757}},
  \href {http://dx.doi.org/10.1016/j.physletb.2009.06.006}
  {\path{doi:10.1016/j.physletb.2009.06.006}}.

\bibitem{Aoki:2006br}
Y.~Aoki, Z.~Fodor, S.~Katz, K.~Szabo, {The QCD transition temperature: Results
  with physical masses in the continuum limit}, Phys. Lett. B643 (2006) 46--54.
\newblock \href {http://arxiv.org/abs/hep-lat/0609068}
  {\path{arXiv:hep-lat/0609068}}, \href
  {http://dx.doi.org/10.1016/j.physletb.2006.10.021}
  {\path{doi:10.1016/j.physletb.2006.10.021}}.

\bibitem{Borsanyi:2010bp}
S.~Borsanyi, et~al., {Is there still any $T_c$ mystery in lattice QCD? Results
  with physical masses in the continuum limit III}, JHEP 1009 (2010) 073.
\newblock \href {http://arxiv.org/abs/1005.3508} {\path{arXiv:1005.3508}},
  \href {http://dx.doi.org/10.1007/JHEP09(2010)073}
  {\path{doi:10.1007/JHEP09(2010)073}}.

\bibitem{Aoki:2006we}
Y.~Aoki, G.~Endrodi, Z.~Fodor, S.~Katz, K.~Szabo, {The Order of the quantum
  chromodynamics transition predicted by the standard model of particle
  physics}, Nature 443 (2006) 675--678.
\newblock \href {http://arxiv.org/abs/hep-lat/0611014}
  {\path{arXiv:hep-lat/0611014}}, \href {http://dx.doi.org/10.1038/nature05120}
  {\path{doi:10.1038/nature05120}}.

\bibitem{Lin:2004en}
Z.-W. Lin, C.~M. Ko, B.-A. Li, B.~Zhang, S.~Pal, {A Multi-phase transport model
  for relativistic heavy ion collisions}, Phys. Rev. C72 (2005) 064901.
\newblock \href {http://arxiv.org/abs/nucl-th/0411110}
  {\path{arXiv:nucl-th/0411110}}, \href
  {http://dx.doi.org/10.1103/PhysRevC.72.064901}
  {\path{doi:10.1103/PhysRevC.72.064901}}.

\bibitem{Peskin:1979va}
M.~E. Peskin, {Short Distance Analysis for Heavy Quark Systems. 1.
  Diagrammatics}, Nucl. Phys. B156 (1979) 365.
\newblock \href {http://dx.doi.org/10.1016/0550-3213(79)90199-8}
  {\path{doi:10.1016/0550-3213(79)90199-8}}.

\bibitem{Manceau:2013zta}
L.~Manceau, {Quarkonium measurements in Pb-Pb and p-Pb collisions with ALICE at
  the LHC}\href {http://arxiv.org/abs/1307.3098} {\path{arXiv:1307.3098}}.

\bibitem{Liu:2012zw}
Y.~Liu, N.~Xu, P.~Zhuang, {Velocity Dependence of Quarkonium Dissociation
  Temperature in High-Energy Nuclear Collisions}, Phys. Lett. B724 (2013)
  73--76.
\newblock \href {http://arxiv.org/abs/1210.7449} {\path{arXiv:1210.7449}},
  \href {http://dx.doi.org/10.1016/j.physletb.2013.05.068}
  {\path{doi:10.1016/j.physletb.2013.05.068}}.

\bibitem{Andronic:2006ky}
A.~Andronic, P.~Braun-Munzinger, K.~Redlich, J.~Stachel, {Statistical
  hadronization of heavy quarks in ultra-relativistic nucleus-nucleus
  collisions}, Nucl. Phys. A789 (2007) 334--356.
\newblock \href {http://arxiv.org/abs/nucl-th/0611023}
  {\path{arXiv:nucl-th/0611023}}, \href
  {http://dx.doi.org/10.1016/j.nuclphysa.2007.02.013}
  {\path{doi:10.1016/j.nuclphysa.2007.02.013}}.

\bibitem{Kostyuk:2005zd}
A.~Kostyuk, {Double, triple and hidden charm production in the statistical
  coalescence model}\href {http://arxiv.org/abs/nucl-th/0502005}
  {\path{arXiv:nucl-th/0502005}}.

\bibitem{Liu:2012tn}
Y.~Liu, C.~Greiner, A.~Kostyuk, {$B_c$ meson enhancement and the momentum
  dependence in Pb+Pb collisions at LHC energy}, Phys. Rev. C87 (2013) 014910.
\newblock \href {http://arxiv.org/abs/1207.2366} {\path{arXiv:1207.2366}},
  \href {http://dx.doi.org/10.1103/PhysRevC.87.014910}
  {\path{doi:10.1103/PhysRevC.87.014910}}.

\bibitem{Grandchamp:2002wp}
L.~Grandchamp, R.~Rapp, {Charmonium suppression and regeneration from SPS to
  RHIC}, Nucl. Phys. A709 (2002) 415--439.
\newblock \href {http://arxiv.org/abs/hep-ph/0205305}
  {\path{arXiv:hep-ph/0205305}}, \href
  {http://dx.doi.org/10.1016/S0375-9474(02)01027-8}
  {\path{doi:10.1016/S0375-9474(02)01027-8}}.

\bibitem{Zhao:2007hh}
X.~Zhao, R.~Rapp, {Transverse Momentum Spectra of $J/\psi$ in Heavy-Ion
  Collisions}, Phys. Lett. B664 (2008) 253--257.
\newblock \href {http://arxiv.org/abs/0712.2407} {\path{arXiv:0712.2407}},
  \href {http://dx.doi.org/10.1016/j.physletb.2008.03.068}
  {\path{doi:10.1016/j.physletb.2008.03.068}}.

\bibitem{Song:2012at}
T.~Song, K.~C. Han, C.~M. Ko, {Charmonium production from nonequilibrium charm
  and anticharm quarks in quark-gluon plasma}, Phys. Rev. C85 (2012) 054905.
\newblock \href {http://arxiv.org/abs/1203.2964} {\path{arXiv:1203.2964}},
  \href {http://dx.doi.org/10.1103/PhysRevC.85.054905}
  {\path{doi:10.1103/PhysRevC.85.054905}}.

\bibitem{Liu:2014rsa}
Y.~Liu, C.~M. Ko, F.~Li, {Heavy quark correlations and the effective volume for
  quarkonia production}\href {http://arxiv.org/abs/1406.6648}
  {\path{arXiv:1406.6648}}.

\bibitem{Abelev:2012kr}
B.~Abelev, et~al., {Inclusive $J/\psi$ production in $pp$ collisions at
  $\sqrt{s} = 2.76$ TeV}, Phys. Lett. B718 (2012) 295--306.
\newblock \href {http://arxiv.org/abs/1203.3641} {\path{arXiv:1203.3641}},
  \href {http://dx.doi.org/10.1016/j.physletb.2012.10.078}
  {\path{doi:10.1016/j.physletb.2012.10.078}}.

\bibitem{Aamodt:2011gj}
K.~Aamodt, et~al., {Rapidity and transverse momentum dependence of inclusive
  J/psi production in $pp$ collisions at $\sqrt{s} = 7$ TeV}, Phys. Lett. B704
  (2011) 442--455.
\newblock \href {http://arxiv.org/abs/1105.0380} {\path{arXiv:1105.0380}},
  \href {http://dx.doi.org/10.1016/j.physletb.2011.09.054,
  10.1016/j.physletb.2012.10.060} {\path{doi:10.1016/j.physletb.2011.09.054,
  10.1016/j.physletb.2012.10.060}}.

\bibitem{Bossu:2011qe}
F.~Bossu, Z.~C. del Valle, A.~de~Falco, M.~Gagliardi, S.~Grigoryan, et~al.,
  {Phenomenological interpolation of the inclusive J/psi cross section to
  proton-proton collisions at 2.76 TeV and 5.5 TeV}\href
  {http://arxiv.org/abs/1103.2394} {\path{arXiv:1103.2394}}.

\bibitem{Liu:2009wza}
Y.~Liu, Z.~Qu, N.~Xu, P.~Zhuang, {Rapidity Dependence of J/psi Production at
  RHIC and LHC}, J. Phys. G37 (2010) 075110.
\newblock \href {http://arxiv.org/abs/0907.2723} {\path{arXiv:0907.2723}},
  \href {http://dx.doi.org/10.1088/0954-3899/37/7/075110}
  {\path{doi:10.1088/0954-3899/37/7/075110}}.

\bibitem{Albacete:2013ei}
J.~L. Albacete, et~al., {Predictions for $p+$Pb Collisions at sqrt s\_NN = 5
  TeV}, Int. J. Mod. Phys. E Vol. 22 (2013) 1330007.
\newblock \href {http://arxiv.org/abs/1301.3395} {\path{arXiv:1301.3395}}.

\bibitem{Wang:2002ck}
X.-N. Wang, F.~Yuan, {Azimuthal asymmetry of J / psi suppression in noncentral
  heavy ion collisions}, Phys. Lett. B540 (2002) 62--67.
\newblock \href {http://arxiv.org/abs/nucl-th/0202018}
  {\path{arXiv:nucl-th/0202018}}, \href
  {http://dx.doi.org/10.1016/S0370-2693(02)02121-4}
  {\path{doi:10.1016/S0370-2693(02)02121-4}}.

\bibitem{Arleo:2001mp}
F.~Arleo, P.~Gossiaux, T.~Gousset, J.~Aichelin, {Heavy quarkonium hadron
  cross-section in QCD at leading twist}, Phys. Rev. D65 (2002) 014005.
\newblock \href {http://arxiv.org/abs/hep-ph/0102095}
  {\path{arXiv:hep-ph/0102095}}, \href
  {http://dx.doi.org/10.1103/PhysRevD.65.014005}
  {\path{doi:10.1103/PhysRevD.65.014005}}.

\bibitem{Zoccoli:2005yn}
A.~Zoccoli, et~al., {Charm, beauty and charmonium production at HERA-B}, Eur.
  Phys. J. C43 (2005) 179--186.
\newblock \href {http://dx.doi.org/10.1140/epjc/s2005-02308-8}
  {\path{doi:10.1140/epjc/s2005-02308-8}}.

\bibitem{Beringer:1900zz}
J.~Beringer, et~al., {Review of Particle Physics (RPP)}, Phys. Rev. D86 (2012)
  010001.
\newblock \href {http://dx.doi.org/10.1103/PhysRevD.86.010001}
  {\path{doi:10.1103/PhysRevD.86.010001}}.

\bibitem{Chen:2013wmr}
B.~Chen, Y.~Liu, K.~Zhou, P.~Zhuang, {$\psi^{\prime}$ Production and B Decay in
  Heavy Ion Collisions at LHC}, Phys. Lett. B726 (2013) 725--728.
\newblock \href {http://arxiv.org/abs/1306.5032} {\path{arXiv:1306.5032}},
  \href {http://dx.doi.org/10.1016/j.physletb.2013.09.036}
  {\path{doi:10.1016/j.physletb.2013.09.036}}.

\end{thebibliography}
\end{document}